\documentclass[11pt]{article}
\usepackage{amssymb,latexsym,amsmath}
\usepackage[dvips]{graphicx}
\usepackage{cite}
\newtheorem{theo}{Theorem}
\newtheorem{defi}[theo]{Definition}
\newtheorem{lemm}[theo]{Lemma}

\newtheorem{prop}[theo]{Proposition}
\def\nn{\nonumber}

\def\R{{\mathbb R}}

\def\su{\mathfrak{su}}
\def\uu{\mathfrak{u}}
\def\so{\mathfrak{so}}

\def\diag{\mathop{\rm diag}\nolimits}
\def\qdots{\mathinner{\mkern1mu\raise1pt\vbox{\kern7pt\hbox{.}}\mkern2mu
 \raise4pt\hbox{.}\mkern2mu\raise7pt\hbox{.}\mkern1mu}}
\newcommand{\myatop}[2]{\genfrac{}{}{0pt}{}{#1}{#2}}
\newcommand\sixj[6]{\left\{ \myatop{#1}{#4} \myatop{#2}{#5} \myatop{#3}{#6}\right\} } % 6j-coeff j1 j2 j3 j4 j5 j6
\def\mybox{\hfill$\Box$}
\begin{document}
\begin{center}
{\Large \bf
The $\su(2)_\alpha$ Hahn oscillator and a discrete Hahn-Fourier transform}\\[5mm]
{\bf E.I.\ Jafarov\footnote{Permanent address: 
Institute of Physics, Azerbaijan National Academy of Sciences, Javid av.\ 33, AZ-1143 Baku, Azerbaijan}, 
N.I.~Stoilova\footnote{Permanent address: 
Institute for Nuclear Research and Nuclear Energy, Boul.\ Tsarigradsko Chaussee 72,
1784 Sofia, Bulgaria} 
and J.\ Van der Jeugt} \\[1mm]
Department of Applied Mathematics and Computer Science,
Ghent University,\\
Krijgslaan 281-S9, B-9000 Gent, Belgium\\[1mm]
E-mail: ejafarov@physics.ab.az, Neli.Stoilova@UGent.be and 
Joris.VanderJeugt@UGent.be
\end{center}

\vskip 10mm
\noindent
Short title: $\su(2)_\alpha$ Hahn oscillator


%\addtolength{\baselineskip}{2mm}
%\addtolength{\abovedisplayskip}{1mm}
%\addtolength{\belowdisplayskip}{1mm}
%\addtolength{\parskip}{1mm}

\begin{abstract}
We define the quadratic algebra $\su(2)_\alpha$ which is a one-parameter deformation of
the Lie algebra $\su(2)$ extended by a parity operator.
The odd-dimensional representations of $\su(2)$ (with representation label $j$, a positive integer)
can be extended to representations of $\su(2)_\alpha$.
We investigate a model of the finite one-dimensional harmonic oscillator based upon this
algebra $\su(2)_\alpha$.
It turns out that in this model the spectrum of the position and momentum operator can be computed
explicitly, and that the corresponding (discrete) wavefunctions can be determined in terms of
Hahn polynomials. 
The operation mapping position wavefunctions into momentum wavefunctions is studied,
and this so-called discrete Hahn-Fourier transform is computed explicitly.
The matrix of this discrete Hahn-Fourier transform has many interesting properties, similar to those of
the traditional discrete Fourier transform.
\end{abstract}

\section{Introduction}

In standard theory of quantum mechanics, position and momentum operators are (essentially) self-adjoint
operators in some infinite-dimensional Hilbert space, satisfying the canonical commutation relations.
Quantum mechanics in finite dimensions has attracted much attention in recent years~\cite{Vourdas}.
In a finite-dimensional Hilbert space, the canonical commutation relations no longer hold.
Despite this, finite-dimensional quantum mechanics has been useful in areas such as quantum computing
and quantum optics~\cite{Braunstein,Miranowicz}.
The defining relations for a quantum mechanical oscillator in a finite-dimensional Hilbert space
are not unique~\cite{Santhanam}, and several type of models have been proposed.
Our interest comes mainly from models related to some Lie algebra (or a deformation thereof).
The finite oscillator model that has been studied most extensively is based on the Lie
algebra $\su(2)$ or $\so(3)$ in the one-dimensional case~\cite{Atak2005,Atak2001,Atak2001b}. 
In the case of a two-dimensional oscillator this has been generalized by the same authors to $\so(4)$.
Such finite oscillator models are of particular importance in optical image processing~\cite{Atak2005},
and more generally in models where only a finite number of eigenmodes can exist.
For example signal analysis dealing with a finite number of discrete sensors or data points led
to physical models realizing a one-dimensional finite oscillator~\cite{Atak1994,Atak1997,Atak1999b}.
In a previous paper~\cite{JSV2011}, a new model for the finite one-dimensional harmonic oscillator was proposed 
based on the algebra $\uu(2)_\alpha$, a one-parameter deformation of the Lie algebra $\uu(2)$. 
This $\uu(2)_\alpha$ model offers an alternative position and momentum spectrum compared to the $\su(2)$ model.
Furthermore, the position wavefunctions have simple expressions in terms of Hahn polynomials,
with interesting properties related to those of a parabose oscillator~\cite{JSV2011}.

For a one-dimensional finite oscillator, one considers three (essentially self-adjoint) operators: 
a position operator $\hat q$, its corresponding momentum operator $\hat p$ and
a (pseudo-) Hamiltonian $\hat H$ which is the generator of time evolution. 
These operators should satisfy the Hamiltonian-Lie equations (or the compatibility of Hamilton's equations with the Heisenberg
equations):
\begin{equation}
[\hat H, \hat q] = -i \hat p, \qquad [\hat H,\hat p] = i \hat q,
\label{Hqp}
\end{equation}
in units with mass and frequency both equal to~1, and $\hbar=1$.
Furthermore, one requires~\cite{Atak2001}:
\begin{itemize}
\item all operators $\hat q$, $\hat p$, $\hat H$ belong to some (Lie) algebra (or superalgebra) $\cal A$;
\item the spectrum of $\hat H$ in (unitary) representations of $\cal A$ is equidistant.
\end{itemize}

The case with ${\cal A}= \su(2)$ (or its enveloping algebra) has been treated considerably in a number of papers~\cite{Atak2001,Atak2001b,Atak2005}.
The relevant representations are the common $\su(2)$ representations labelled by an integer or half-integer $j$.
In such a representation, the Hamiltonian is taken as $\hat H=J_0+j+\frac12$, where $J_0=J_z$ is the diagonal $\su(2)$
operator. Thus the spectrum of $\hat H$ is $n+\frac12$ ($n=0,1,\ldots,2j$).
With $\hat q = \frac12(J_++J_-)=J_x$ and $\hat p = \frac{i}{2}(J_+-J_-)=-J_y$, the relations~\eqref{Hqp} 
are satisfied. Clearly, $\hat q$ and $\hat p$ have a finite spectrum given by $\{-j,-j+1,\ldots,+j\}$~\cite{Atak2001}.
More important, the position wavefunctions have been constructed, and are given by Krawtchouk functions (normalized
symmetric Krawtchouk polynomials). 
These discrete wavefunctions have interesting properties, and their shape is reminiscent of those of the canonical oscillator~\cite{Atak2001}. This is explained by the fact that under the limit $j\rightarrow \infty$ 
the discrete wavefunctions coincide with the canonical wavefunctions in terms of Hermite polynomials~\cite{Atak2001,Atak2003}.

In~\cite{JSV2011}, the case with ${\cal A}=\uu(2)_\alpha$ was investigated ($\alpha>-1$).
For this one-parameter deformation of $\uu(2)$, only the representations labelled by a half-integer $j$ survive as
representations of $\uu(2)_\alpha$ (so only the even-dimensional representations).
This led to an alternative model of the finite oscillator, with the spectrum of $\hat H$ 
again given by $n+\frac12$ ($n=0,1,\ldots,2j$).
For the operators $\hat q$ and $\hat p$, the spectrum is again finite and equidistant in steps of one unit, 
except that there is a gap of size $2\alpha+2$ in the middle; explicitly, it is given by
\begin{equation}
-\alpha-j-\frac12, -\alpha-j+\frac12, \ldots, -\alpha-1; \alpha+1,\alpha+2, \ldots,\alpha+j+\frac12.
\label{q-spectrum-even}
\end{equation}
An interesting result in~\cite{JSV2011} was that the position wavefunctions could be constructed and
they turned out to be normalized Hahn (or dual Hahn) polynomials. 
An investigation of these discrete wavefunctions gave rise to remarkable plots, suggesting in fact a
relation with the parabose oscillator. Indeed it was shown that under the limit $j\rightarrow \infty$ 
the discrete wavefunctions coincide with the parabose wavefunctions in terms of Laguerre polynomials~\cite{JSV2011}.

Despite the novel results presented in~\cite{JSV2011}, it remains somehow intriguing that the 
odd-dimen\-sion\-al $\uu(2)$ representations can not be deformed as $\uu(2)_\alpha$ representations.
In the present paper, we present the solution to this problem.
It turns out that one should consider a different one-parameter deformation of $\su(2)$, 
involving only the three $\su(2)$ operators and a parity operator (and no central element).
This new deformation, denoted by $\su(2)_\alpha$, is defined and it is shown that in this case
the odd-dimensional $\su(2)$ representations, labelled by an integer value $j$, can be
deformed as $\su(2)_\alpha$ representations. 
Using $\su(2)_\alpha$ as a model for the finite oscillator, yields the same spectrum of $\hat H$ 
given by $n+\frac12$ ($n=0,1,\ldots,2j$). The finite spectrum of the position and momentum
operators $\hat q$ and $\hat p$ is however quite different, and given by
\[
0, \quad \pm \sqrt{k(2\alpha+k+1)}, \qquad (k=1,\ldots,j).
\]
The position (and momentum) wavefunctions are constructed, and turn out to be again Hahn
polynomials (in this case with parameters $(\alpha,\alpha)$ or $(\alpha+1,\alpha+1)$).
These wavefunctions are once more discrete versions of the parabose wavefunctions, as a limit
computation shows.
A fascinating question in this context is the relation between the position wavefunctions and the
momentum wavefunctions. In the canonical case, these functions are related through the Fourier transform.
Here, we show that these wavefunctions are related through a discrete version of the Fourier
transform, which we refer to as the discrete Hahn-Fourier transform.
For readers primarily interested in special functions, the computation of this discrete Hahn-Fourier transform
is probably the most attractive part. It involves a special case of a bilinear generating
function (the Poisson kernel) for dual Hahn polynomials.

The contents of the remaining sections is as follows: in section~2 the deformed algebra $\su(2)_\alpha$ and its
representations are constructed. In section~3 we use $\su(2)_\alpha$ as a model for the finite oscillator,
and determine in particular the spectrum of the position and momentum operators, and their eigenvectors.
The structure of these eigenvectors is studied in section~4, yielding position and momentum wavefunctions.
In section~5 we determine the operation that transforms position wavefunction into momentum wavefunctions,
the so-called discrete Hahn-Fourier transform $F$, and its properties. 
The somewhat technical but interesting proof of the ${}_4F_3(1)$ form of the matrix elements of $F$ is given
in section~6. Finally we give a summary of the results in section~7.

\section{The algebra $\su(2)_\alpha$ and its representations}

The Lie algebra $\su(2)$~\cite{Wybourne,Humphreys} can be defined by its basis elements 
$J_0$, $J_+$, $J_-$ with commutators $[J_0,J_\pm]=\pm J_\pm$ and $[J_+,J_-]=2J_0$.
The non-trivial unitary representations of $\su(2)$, corresponding to the star relations $J_0^\dagger=J_0$, 
$J_\pm^\dagger=J_\mp$, are labelled~\cite{Wybourne,Humphreys} by a positive integer or half-integer $j$ and
have dimension $2j+1$. The action on a standard basis vectors $|j,m\rangle$ (with $m=-j,-j+1,\ldots,+j$) is given by
\[
J_0 |j,m\rangle = m\;|j,m\rangle,\qquad 
J_\pm |j,m\rangle = \sqrt{(j\mp m)(j\pm m +1)}\;|j,m\pm 1\rangle.
\]
The universal enveloping algebra of $\su(2)$ can be extended by a parity operator $P$ with action
$P |j,m\rangle = (-1)^{j+m}\;|j,m\rangle$. This means 
that $P$ commutes with $J_0$, anticommutes with $J_+$ and $J_-$, and $P^2=1$.
This extended algebra can be deformed by a parameter $\alpha$, leading to the definition of $\su(2)_\alpha$.

\begin{defi}
Let $\alpha$ be a parameter. The algebra $\su(2)_\alpha$ is a unital algebra with basis elements 
$J_0$, $J_+$, $J_-$ and $P$ subject to the following relations:
\begin{itemize}
\item $P$ is a parity operator satisfying $P^2=1$ and
\begin{equation}
[P,J_0]=PJ_0-J_0P=0, \qquad \{P,J_\pm\}=PJ_\pm + J_\pm P= 0.
\label{P}
\end{equation}
\item The $\su(2)$ relations are deformed as follows:
\begin{align}
& [J_0, J_\pm] = \pm J_\pm,  \label{J0J+} \\
& [J_+, J_-] = 2 J_0 + 2(2\alpha+1)J_0P.
\label{J+J-}
\end{align}
\end{itemize}
\end{defi}

Note that this is different from the $\uu(2)_\alpha$ deformation introduced in~\cite{JSV2011}; in particular it does
not involve a central element. For the deformation $\uu(2)_\alpha$, the even-dimensional $\uu(2)$ representations
($j$ half-integer) could be deformed. 
The current $\su(2)_\alpha$ is ``complementary'' in the sense that now the odd-dimensional
$\su(2)$ representations ($j$ integer) can be deformed:

\begin{prop}
Let $j$ be an integer (i.e.\ $2j$ is even), and consider the space $W_j$ with basis vectors 
$|j,-j\rangle$, $|j,-j+1\rangle$, $\ldots$, $|j,j\rangle$. Assume that $\alpha>-1$. Then the following action turns $W_j$ into
an irreducible representation space of $\su(2)_\alpha$.
\begin{align}
& P |j,m\rangle = (-1)^{j+m}\;|j,m\rangle,\label{act-P}\\
& J_0 |j,m\rangle = m\;|j,m\rangle,\label{act-J0}\\
& J_+ |j,m\rangle = 
  \begin{cases}
 \sqrt{(j-m)(j+m +2\alpha +2)}\;|j,m+1\rangle, & \text{if $j+m$ is even;}\\
 \sqrt{(j-m+2\alpha+1)(j+m+1)}\;|j,m+1\rangle, & \text{if $j+m$ is odd,}
 \end{cases} \label{act-J+}\\
& J_- |j,m\rangle = 
  \begin{cases}
 \sqrt{(j+m +2\alpha +1)(j-m+1)}\;|j,m-1\rangle, & \text{if $j+m$ is odd;}\\
 \sqrt{(j+m)(j-m+2\alpha+2)}\;|j,m-1\rangle, & \text{if $j+m$ is even.}
 \end{cases} \label{act-J-}
\end{align} 
\end{prop}

The proof is essentially by direct computation, the same as~\cite[Proposition~2]{JSV2011}.
Note that the representation given in this proposition is unitary under the star conditions  
$P^\dagger=P$, $J_0^\dagger=J_0$, $J_\pm^\dagger=J_\mp$.
Also note that for $\alpha=-\frac12$, the deformation is trivial (both for the algebra and the representations).

\section{Using $\su(2)_\alpha$ as a model for the one-dimensional oscillator}

Quite similar as in the non-deformed case~\cite{Atak2005} or in the $\uu(2)_\alpha$ deformed case~\cite{JSV2011}, 
let us choose the position, momentum and Hamiltonian (representation dependent) operators as follows:
\begin{equation}
\hat q = \frac12 (J_++J_-), \qquad
\hat p = \frac{i}{2}(J_+-J_-), \qquad
\hat H = J_0+j+\frac12 .
\end{equation}
These operators satisfy~\eqref{Hqp}. 
In the representation space $W_j$, $\hat H |j,m\rangle = (m+j+\frac12)|j,m\rangle$, 
therefore the spectrum of $\hat H$ is linear and given by
\begin{equation}
n+\frac12 \qquad (n=0,1,\ldots,2j).
\end{equation}
{}From the actions~\eqref{act-J+}-\eqref{act-J-}, one finds
\begin{equation*}
2\hat q |j,m\rangle = \sqrt{(j+m)(j-m +2\alpha +2)}\;|j,m-1\rangle +
\sqrt{(j-m)(j+m+2\alpha+2)}\;|j,m+1\rangle,
\end{equation*}
if $j+m$ is even, and
\begin{equation*}
2\hat q |j,m\rangle = \sqrt{(j-m+1)(j+m+2\alpha+1)}\;|j,m-1\rangle +
\sqrt{(j+m+1)(j-m +2\alpha +1)}\;|j,m+1\rangle,
\end{equation*}
if $j+m$ is odd. The action of $2i\hat p$ is similar. 
In the (ordered) basis $\{ |j,-j\rangle, |j,-j+1\rangle, \ldots, |j,j-1\rangle, |j,j\rangle \}$ of $W_j$,
the operators $2\hat q$ and $2i\hat p$ take the matrix form
\begin{align}
2\hat q&=\left(
\begin{array}{ccccc}
0 & M_0& 0 & \cdots & 0 \\
M_0 & 0 & M_1 & \cdots & 0\\
0 & M_1 & 0 & \ddots &  \\
\vdots & \vdots & \ddots & \ddots& M_{2j-1}\\
0 & 0 &  & M_{2j-1} & 0
\end{array}
\right)\equiv M^q,
\label{Mq} \\
2i\hat p&=\left(
\begin{array}{ccccc}
0 & M_0& 0 & \cdots & 0 \\
-M_0 & 0 & M_1 & \cdots & 0\\
0 & -M_1 & 0 & \ddots &  \\
\vdots & \vdots & \ddots & \ddots& M_{2j-1}\\
0 & 0 &  & -M_{2j-1} & 0
\end{array}
\right)\equiv M^p,
\label{Mp}
\end{align}
with
\begin{equation}
M_k= \begin{cases}
 \sqrt{(k+1)(2j+2\alpha -k+1)}, & \text{if $k$ is odd;}\\
 \sqrt{(k+2\alpha+2)(2j-k)}, & \text{if $k$ is even.}
 \end{cases}
\label{Ma}
\end{equation}
For these matrices the eigenvalues and eigenvectors can be constructed in terms of Hahn polynomials.
Hahn polynomials $Q_n(x;\alpha, \beta, N)$~\cite{Koekoek,Ismail} of degree $n$ ($n=0,1,\ldots,N$) in the variable $x$, 
with parameters $\alpha>-1$ and $\beta>-1$ are defined by~\cite{Koekoek,Ismail}:
\begin{equation}
Q_n(x;\alpha,\beta,N) = {\;}_3F_2 \left( \myatop{-n,n+\alpha+\beta+1,-x}{\alpha+1,-N} ; 1 \right),
\label{defQ}
\end{equation}
in terms of the generalized hypergeometric series $_3F_2$ of unit argument~\cite{Bailey,Slater}. 
Their (discrete) orthogonality relation reads~\cite{Koekoek,Ismail}:
\begin{equation}
\sum_{x=0}^N w(x;\alpha, \beta,N) Q_l(x;\alpha, \beta, N) Q_n(x;\alpha,\beta,N) = h(n;\alpha,\beta,N)\, \delta_{ln},
\label{orth-Q}
\end{equation} 
where
\begin{align*}
& w(x;\alpha, \beta,N) = \binom{\alpha+x}{x} \binom{N+\beta-x}{N-x} \qquad (x=0,1,\ldots,N); \\
& h(n;\alpha,\beta,N)= \frac{(n+\alpha+\beta+1)_{N+1}(\beta+1)_n n!}{(2n+\alpha+\beta+1)(\alpha+1)_n(N-n+1)_n N!}.
\end{align*}
We have used here the common notation for Pochhammer symbols~\cite{Bailey,Slater}
$(a)_k=a(a+1)\cdots(a+k-1)$ for $k=1,2,\ldots$ and $(a)_0=1$.
Orthonormal Hahn functions $\tilde Q$ are determined by:
\begin{equation}
\tilde Q_n(x;\alpha,\beta,N) \equiv \frac{\sqrt{w(x;\alpha,\beta,N)}\, Q_n(x;\alpha,\beta,N)}{\sqrt{h(n;\alpha,\beta,N)}}.
\label{Q-tilde}
\end{equation}
Recall that dual Hahn polynomials have a similar expression. In fact, for $x\in\{0,1,\ldots,N\}$ the right hand
side of~\eqref{defQ} is the dual Hahn polynomial $R_x(\lambda(n);\alpha,\beta,N)$ of degree $x$ in the variable
$\lambda(n)=n(n+\alpha+\beta+1)$; see~\cite{Koekoek,Ismail} for their orthogonality relations.

In terms of these, the eigenvalues and (orthonormal) eigenvectors of $M^q$ can be constructed:
\begin{prop}
Let $M^q\equiv 2\hat q$ be the tridiagonal $(2j+1)\times(2j+1)$-matrix~\eqref{Mq} and let $U=(U_{kl})_{0\leq k,l\leq 2j}$
be the $(2j+1)\times(2j+1)$-matrix with matrix elements:
\begin{align}
& U_{2i,j} =(-1)^i \tilde Q_0(i;\alpha,\alpha,j), \;i\in\{0,1,\ldots,j\};\;
U_{2i+1,j} = 0, \; i\in\{0,\ldots,j-1\}; \label{Uj}\\
& U_{2i,j-k} = U_{2i,j+k} = \frac{(-1)^i}{\sqrt{2}} \tilde Q_k(i;\alpha,\alpha,j)
, \;i\in\{0,1,\ldots,j\}, \; k\in\{1,\ldots,j\};  \label{Ueven}\\
& U_{2i+1,j-k} = -U_{2i+1,j+k} = -\frac{(-1)^i}{\sqrt{2}} \tilde Q_{k-1}(i;\alpha+1,\alpha +1,j-1), \; 
i\in\{0,1,\ldots,j-1\}, \nn \\
& \hskip 10.5 cm k \in\{1,\ldots,j\}\label{Uodd}.
\end{align}
Then $U$ is an orthogonal matrix:
\begin{equation}
U U^T = U^TU=I,
\end{equation}
the columns of $U$ are the eigenvectors of $M^q$, i.e.
\begin{equation}
M^q U = U D^q,
\label{MUUD}
\end{equation}
where $D^q= \diag (\epsilon_0,\epsilon_1,\ldots,\epsilon_{2j})$ is a diagonal matrix 
containing the eigenvalues $\epsilon_k$ of $M^q$:
\begin{equation}
\epsilon_{j-k}=-2\sqrt{k(2\alpha +k+1)}, \quad \epsilon_{j}=0, \quad \epsilon_{j+k}=2\sqrt{k(2\alpha+k+1)}, \label{epsilon} 
\quad (k=1,\ldots,j).
\end{equation}
\label{propU}
\end{prop}

\noindent {\bf Proof.}
Using the orthogonality of the Hahn polynomials, and the explicit expressions~\eqref{Uj}-\eqref{Uodd}, 
a simple computation shows that $(U^TU)_{kl}=\delta_{kl}$. Thus $U^TU=I$, the identity matrix, and 
hence $UU^T=I$ holds as well.

Now  it remains to verify~\eqref{MUUD} and~\eqref{epsilon}. By the form of $M^q$:
\begin{equation}
\big(M^qU\big)_{ik}= \sum_{l=0}^{2j}M_{il}^q U_{lk}=M_{i-1}U_{i-1,k}+M_{i}U_{i+1,k}.
\label{MU}
\end{equation}
We have to consider \eqref{MU} in six distinct cases, according to $i$ even or odd, and to $k$ belonging to $\{0,1,\ldots,j-1\}$, to $\{j+1,j+2,\ldots,2j\}$ or $k=j$. 
Let us consider the case that $i$ is odd and $k\in\{0,1,\ldots,j-1\}$. Then, relabelling the indices appropriately, \eqref{Ma}, \eqref{Ueven} and~\eqref{Uodd} yield:
\begin{align*}
&(M^qU)_{2i+1,j-k}=M_{2i}U_{2i,j-k}+M_{2i+1}U_{2i+2,j-k}  \\
& =(-1)^i\sqrt{2}\sqrt{(\alpha+i+1)(j-i)}\tilde Q_k(i;\alpha,\alpha,j)+
(-1)^{i+1}\sqrt{2}\sqrt{(i+1)(j+\alpha-i)}\tilde Q_k(i+1;\alpha,\alpha,j)\\
&= (-1)^i \sqrt{2}\sqrt{\frac{(\alpha+1)_{i+1}(\alpha+1)_{j-i}}{i!(j-i-1)!h(k;\alpha,\alpha,j)}}[  Q_k(i;\alpha,\alpha,j)- Q_k(i+1;\alpha,\alpha,j)].
\end{align*}
At this point, the forward shift operator formula for Hahn polynomials~\cite[(9.5.6)]{Koekoek} can be applied and
yields
\begin{align*}
(M^qU)_{2i+1,j-k}& =(-1)^i \sqrt{2}  \sqrt{\frac{(\alpha+1)_{i+1}(\alpha+1)_{j-i}}{i!(j-i-1)!h(k;\alpha,\alpha,j)}} \frac{k(k+2\alpha+1)}{(\alpha+1)j} Q_{k-1}(i;\alpha+1,\alpha+1,j-1)\\
&= -2\sqrt{k(k+2\alpha+1)} U_{2i+1,j-k} = \epsilon_{j-k}U_{2i+1,j-k}=\big(UD^q\big)_{2i+1,j-k}.
\end{align*}
For the other five cases, the computations are similar and for some of them the backward shift operator formula for Hahn polynomials~\cite[(9.5.8)]{Koekoek} is applied. 
\mybox

\begin{figure}[htb]
\begin{center}
\begin{tabular}{lc}
(a) &\includegraphics[scale=1.5]{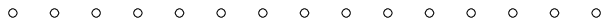}  \\[5mm]
(b) &\includegraphics[scale=1.5]{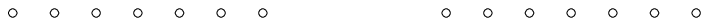}  \\[5mm]
(c) &\includegraphics[scale=1.5]{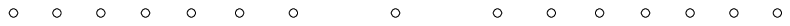}  
\end{tabular} 
\end{center}
\caption{Plots of a typical spectrum of the position operator, (a) in the case of the $\su(2)$ model,
(b) in the case of the $\uu(2)_\alpha$ model, and (c) in the case of the $\su(2)_\alpha$ model.}
\label{fig1}
\end{figure}

Note that in the case of $\uu(2)_\alpha$~\cite{JSV2011}, the equation corresponding to~\eqref{MU} was
related to two new difference equations for Hahn polynomials~\cite{SV2011,Gorin}.
Here, the equation is just corresponding to known forward or backward shift operator formulas.
The above proposition yields in particular the spectrum of the position operator:
\begin{prop}
The $2j+1$ eigenvalues $q$ of the position operator $\hat q$ in the representation $W_j$ are given by
\begin{equation}
-\sqrt{j(2\alpha+j+1)}, -\sqrt{(j-1)(2\alpha+j)}, \ldots,  -\sqrt{2\alpha+2}; 0 ; 
\sqrt{2\alpha+2},  \ldots,  \sqrt{j(2\alpha+j+1)}.
\end{equation}
It will be appropriate to label these $\hat q$-eigenvalues as $q_k$, where $k=-j,-j+1,\ldots,+j$, so
\[
q_{\pm k} = \pm\sqrt{k(2\alpha+k+1)}, \qquad k=0, 1, \ldots, j.
\]
\label{prop4}
\end{prop}
One can compare the spectrum of $\hat q$ with that in related models, see Figure~\ref{fig1}.
In the non-deformed case ($\alpha=-\frac12$, or the $\su(2)$ model), the spectrum is just equidistant as
already mentioned in the introduction.
In the case of $\uu(2)_\alpha$ (and $j$ half-integer), the spectrum is given by~\eqref{q-spectrum-even}: apart from an extra gap
of size $2\alpha+2$ in the middle, it is again equidistant.
In the current case of $\su(2)_\alpha$ (and $j$ integer), the spectrum given by Proposition~\ref{prop4}
is not equidistant.

Essentially the eigenvectors of~\eqref{Mq} have components proportional to Hahn polynomials
with parameters $(\alpha,\alpha)$ when the component has even index and with parameters ($\alpha+1,\alpha+1)$
when the component has odd index.
It is convenient to introduce a notation for these eigenvectors:
the orthonormal eigenvector of the position operator $\hat q$ in $W_j$ for the eigenvalue $q_k$, denoted by $|j,q_k)$, is given 
in terms of the standard basis by
\begin{equation}
|j,q_k) = \sum_{m=-j}^j U_{j+m,j+k} |j,m\rangle.
\end{equation}

Let us now turn our attention to the momentum operator $\hat p$.
Up to signs, the matrix $M^p$ is the same as the matrix $M^q$. The analysis of its eigenvalues and eigenvectors
is thus very similar. We present just the final result here:

\begin{prop}
Let $M^p\equiv 2i\hat p$ be the tridiagonal $(2j+1)\times(2j+1)$-matrix~\eqref{Mp} and let $V=(V_{kl})_{0\leq k,l\leq 2j}$
be the $(2j+1)\times(2j+1)$-matrix with matrix elements
\begin{equation}
V_{k,l}= - i^{k+1} U_{kl},
%V_{2k,l}=-i(-1)^k U_{2k,l},\qquad V_{2k+1,l}= (-1)^k U_{2k+1,l},
\label{V}
\end{equation}
where $U$ is the matrix determined by \eqref{Uj}-\eqref{Uodd}.
Then $V$ is a unitary matrix, $V V^\dagger = V^\dagger V=I$.
The columns of $V$ are the eigenvectors of $M^p$, i.e.
\begin{equation}
M^p V = V D^p,
\label{MVVD}
\end{equation}
where $D^p$ is a diagonal matrix containing the eigenvalues $\varepsilon_k$ of $M^p$:
\begin{align}
& D^p= \diag (\varepsilon_0,\varepsilon_1,\ldots,\varepsilon_{2j}), \nn\\[1mm]
& \varepsilon_{j-k}=-2i\sqrt{k(2\alpha +k+1)}, \quad \epsilon_{j}=0, \quad \epsilon_{j+k}=2i\sqrt{k(2\alpha+k+1)}  
\quad (k=1,\ldots,j). \label{varepsilon}
\end{align}
\end{prop}

Hence the $2j+1$ eigenvalues $p$ of the momentum operator $\hat p$ in the representation $W_j$ are given by
\begin{equation}
-\sqrt{j(2\alpha+j+1)}, -\sqrt{(j-1)(2\alpha+j)}, \ldots,  -\sqrt{2\alpha+2}; 0 ; 
\sqrt{2\alpha+2},  \ldots,  \sqrt{j(2\alpha+j+1)},
\end{equation}
in other words, the momentum operator $\hat p$ has the same spectrum as the position operator $\hat q$.
We shall denote these $\hat p$-eigenvalues by $p_k$, where $k=-j,-j+1,\ldots,+j$, so
\[
p_{\pm k} = \pm\sqrt{k(2\alpha+k+1)}, \qquad k=0, 1, \ldots, j.
\]
The normalized eigenvector of the momentum operator $\hat p$ in $W_j$ for the eigenvalue $p_k$, 
denoted by $|j,p_k)$, is then given by
\begin{equation}
|j,p_k) = \sum_{m=-j}^j V_{j+m,j+k} |j,m\rangle.
\end{equation}

Finally, it is worth mentioning another property of the matrix elements of $V$, that follows from the explicit
expressions~\eqref{V}, \eqref{Uj}-\eqref{Uodd} and the orthogonality properties of the Hahn polynomials:
\begin{equation}
V^T V = \left( \begin{array}{cccc}
0 & \cdots & 0 & -1 \\ 0 & \cdots & -1 & 0 \\ \vdots & \qdots & \vdots &\vdots\\ -1 & \cdots & 0 & 0 \\ \end{array}\right).
\label{VTV}
\end{equation}
Note also that from~\eqref{V} it follows that
\begin{equation}
V={\cal J}U \qquad\hbox{where}\qquad
{\cal J}=\diag(-i,1,i,-1,\ldots),
\label{VJU}
\end{equation}
and the sequence $(-i,1,i,-1)$ is repeated, ending with $i$ or $-i$ (since the matrices are odd-dimensional).
One can also write ${\cal J}= -i \diag(i^0,i^1,i^2,i^2,\ldots,i^{2j})$.

\section{Position and momentum wavefunctions and their properties}

The position (resp.\ momentum) wavefunctions of the $\su(2)_\alpha$ finite oscillator are the overlaps 
between the $\hat q$-eigenvectors (resp.\ $\hat p$-eigenvectors)
and the $\hat H$-eigenvectors (or equivalently, the $J_0$-eigenvectors $|j,m\rangle$).
Let us denote them by $\Phi^{(\alpha)}_{j+m}(q)$ (resp.\ $\Psi^{(\alpha)}_{j+m}(p)$ ), where $m=-j,-j+1,\ldots,+j$, and where $q$ (resp.\ $p$) assumes one of the 
discrete values $q_k$ (resp.\ $p_k$) $(k=-j,-j+1,\ldots,+j)$. 
Therefore, in the notation of the previous section:
\begin{align}
& \Phi^{(\alpha)}_{j+m}(q_k)= \langle j,m | j,q_k ) = U_{j+m,j+k}, \label{UPhi}\\
& \Psi^{(\alpha)}_{j+m}(p_k)= \langle j,m | j,p_k ) = V_{j+m,j+k}. \label{VPsi}
\end{align}
Let us consider the explicit form of these wavefunctions, first for the position variable.
For $j+m$ even, $j+m=2n$, and for positive $q$-values one has
\begin{equation}
\Phi^{(\alpha)}_{2n} (q_{k}) = \frac{(-1)^n}{\sqrt{2}} \tilde Q_{k}(n;\alpha,\alpha,j), \qquad n=0, 1, \ldots, j, 
\qquad k=1, \ldots, j,
\end{equation}
or equivalently:
\begin{align}
\Phi^{(\alpha)}_{2n} (q_{k}) &= \frac{(-1)^n}{\sqrt{2}} 
\sqrt{\frac{w(n;\alpha,\alpha,j)}{h(k;\alpha,\alpha,j)}} 
{\ }_3F_2 \left( \myatop{-k,k+2\alpha+1,-n}{\alpha+1,-j} ; 1 \right)\nn \\
& =\frac{(-1)^n}{\sqrt{2}} 
\sqrt{\frac{w(n;\alpha,\alpha,j)}{h(k;\alpha,\alpha,j)}}  
R_n \left( q_k^2; \alpha, \alpha, j\right), \qquad q_{k}^2 = k(2\alpha+k+1),
\label{Phi-even}
\end{align}
where $R_n \left( \lambda_k; \alpha, \alpha, j\right)$ is a dual Hahn polynomial~\cite{Koekoek,Ismail} 
of degree $n$ in the variable $\lambda(k)\equiv q_k^2=k(2\alpha+k+1)$.
In a similar way, one finds for $j+m$ odd, $j+m=2n+1$, and for positive $q$-values, that
\begin{align}
\Phi^{(\alpha)}_{2n+1} (q_{k}) & = \frac{(-1)^n}{\sqrt{2}} 
\sqrt{\frac{w(n;\alpha+1,\alpha+1,j-1)}{h(k-1;\alpha+1,\alpha+1,j-1)}} 
{\ }_3F_2 \left( \myatop{-k+1,k+2\alpha+2,-n}{\alpha+2,-j+1} ; 1 \right)\nn \\
&=\sqrt{\frac{w(n;\alpha+1,\alpha+1,j-1)}{h(k-1;\alpha+1,\alpha+1,j-1)}} 
R_n \left( q_k^2-2(\alpha+1);\alpha+1,\alpha+1,j-1 \right).
\label{Phi-odd}
\end{align}
For $q=0$ or negative $q$-values, the expressions are of course analogous, according to~\eqref{Uj}-\eqref{Uodd}.

\begin{figure}[htb]
\begin{tabular}{cc}
\hline\\[-3mm]
\includegraphics[scale=0.55]{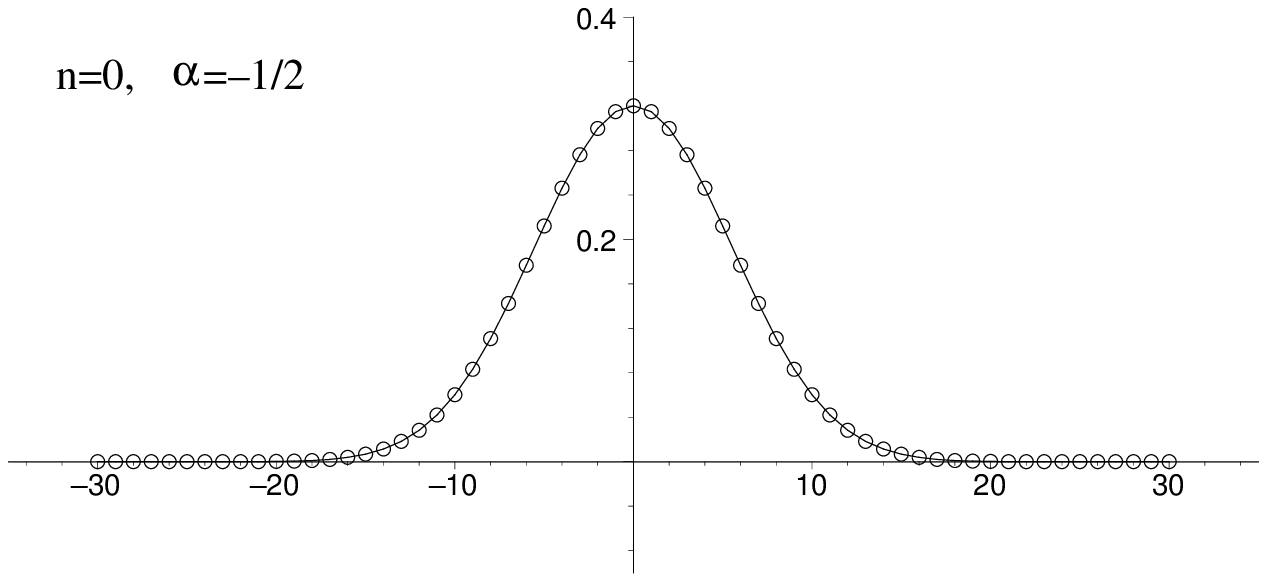} & \includegraphics[scale=0.55]{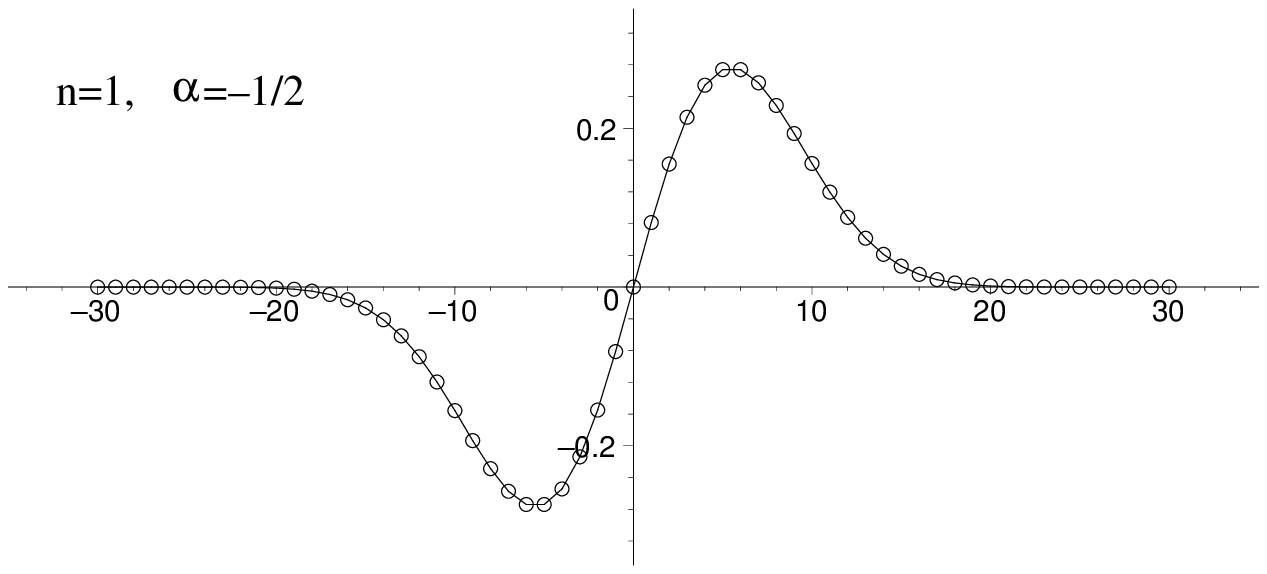} \\
\includegraphics[scale=0.55]{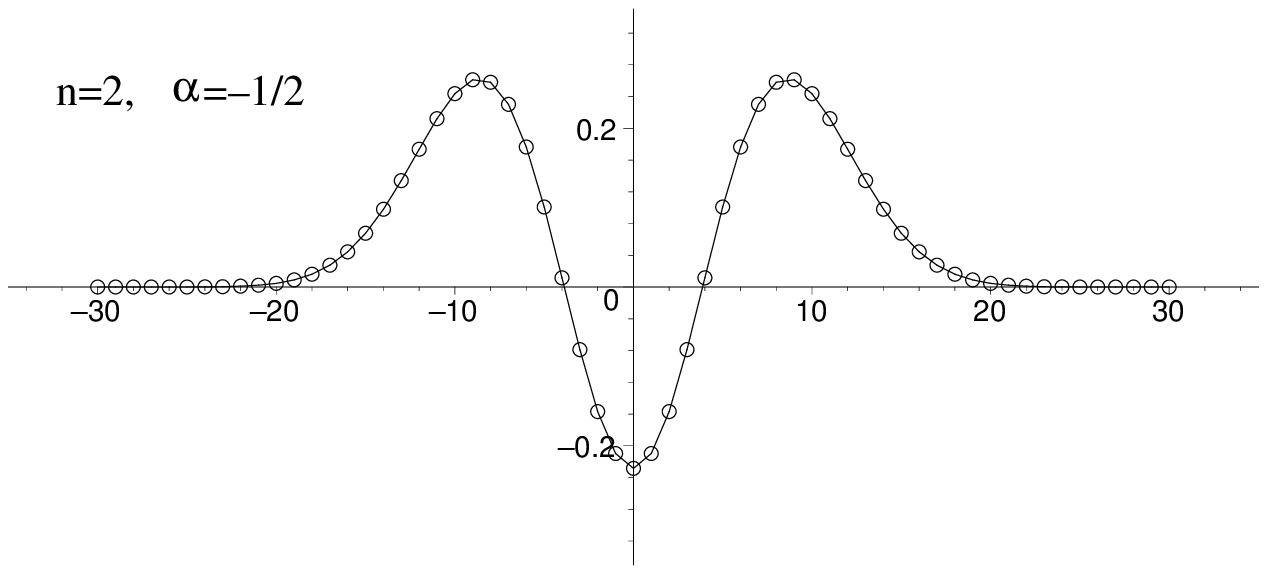} & \includegraphics[scale=0.55]{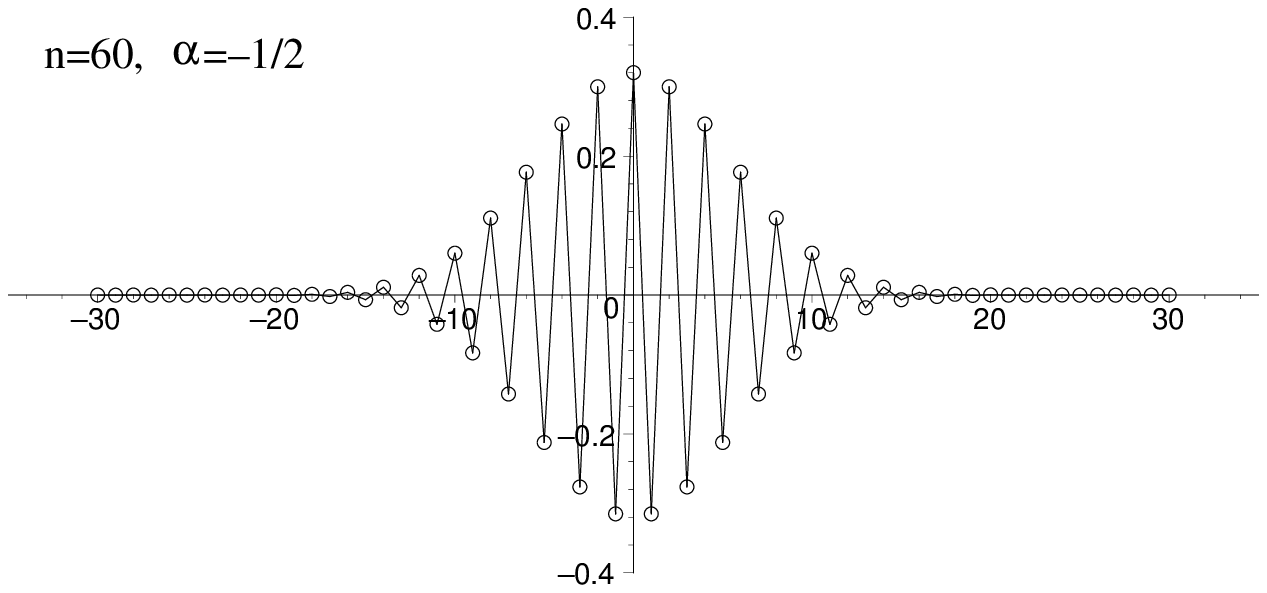} \\[-1mm]
\hline\\[-3mm]
\includegraphics[scale=0.55]{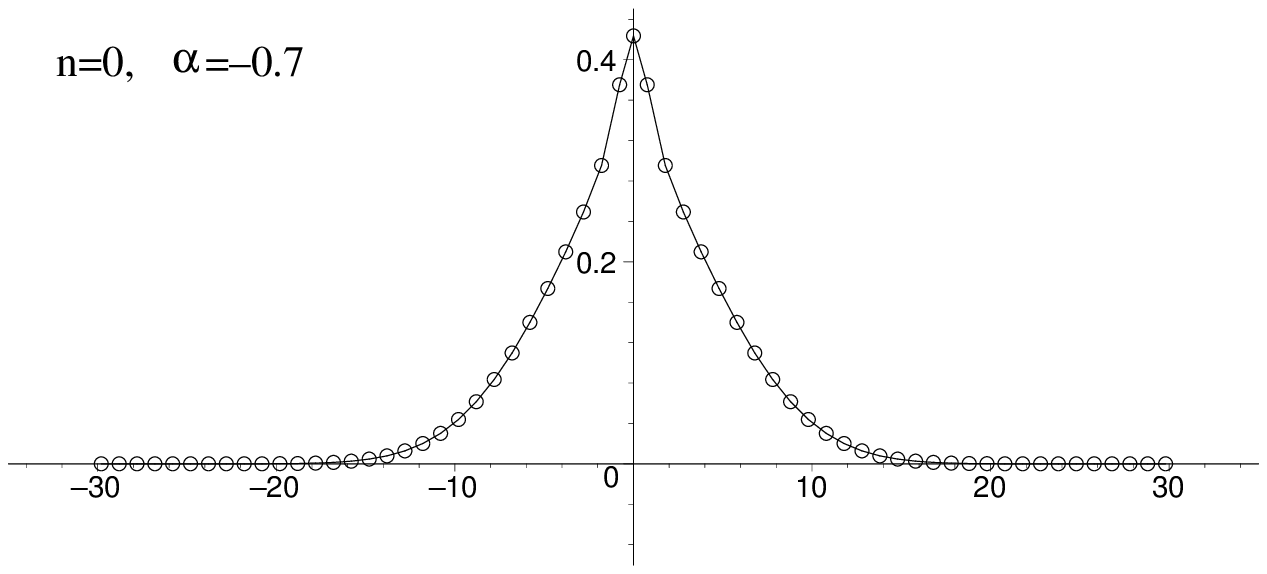} & \includegraphics[scale=0.55]{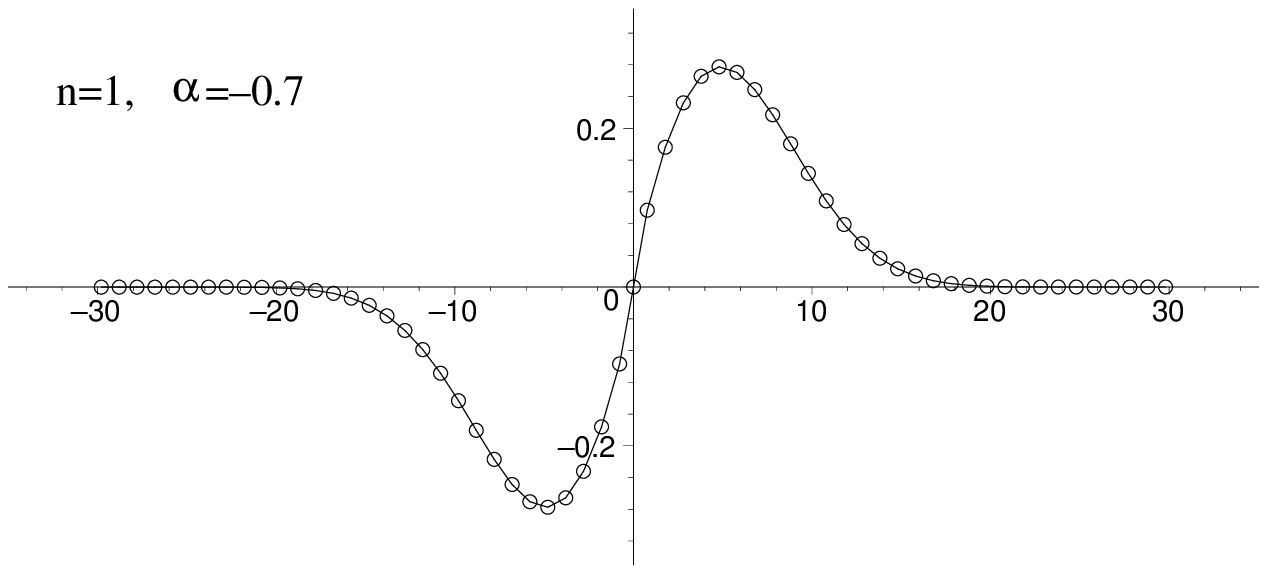} \\
\includegraphics[scale=0.55]{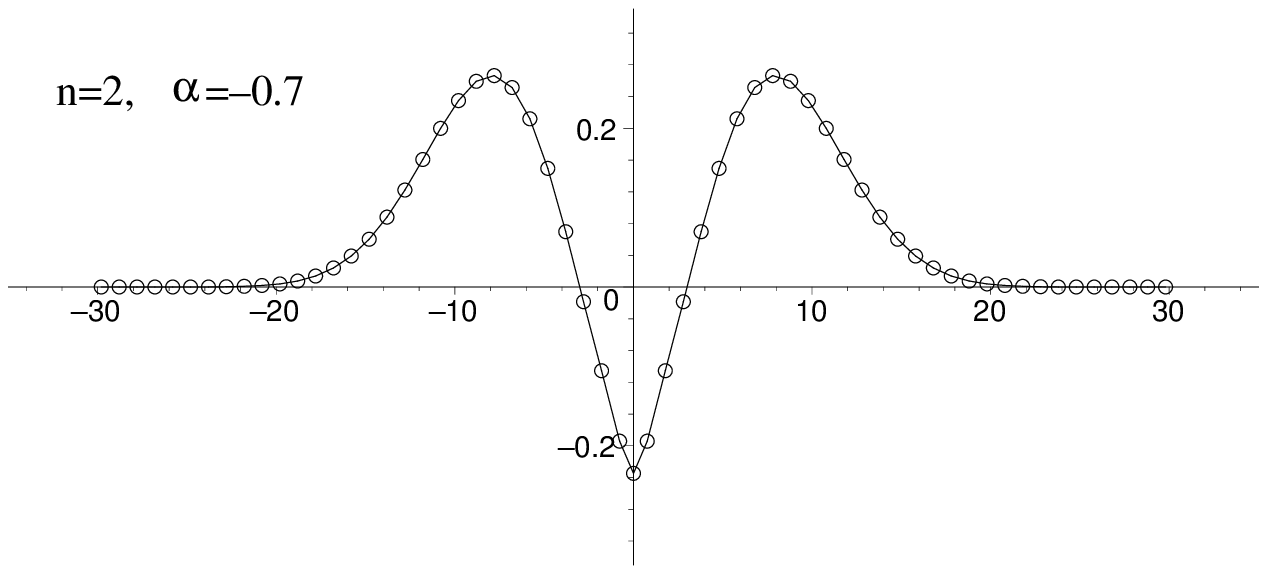} & \includegraphics[scale=0.55]{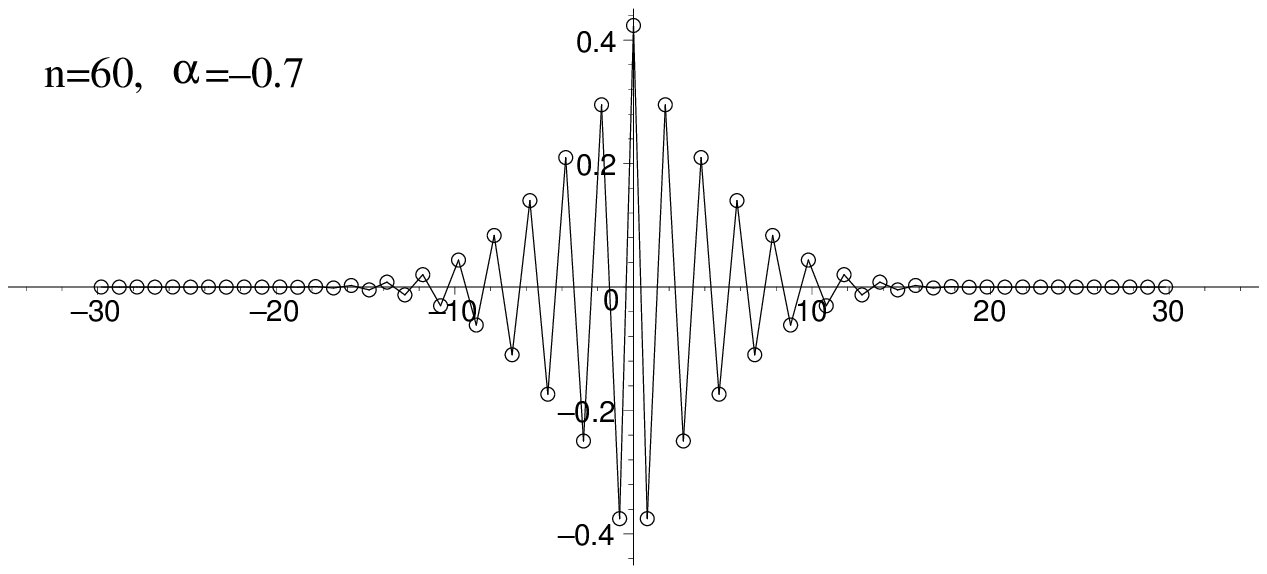} \\[-1mm]
\hline\\[-3mm]
\includegraphics[scale=0.55]{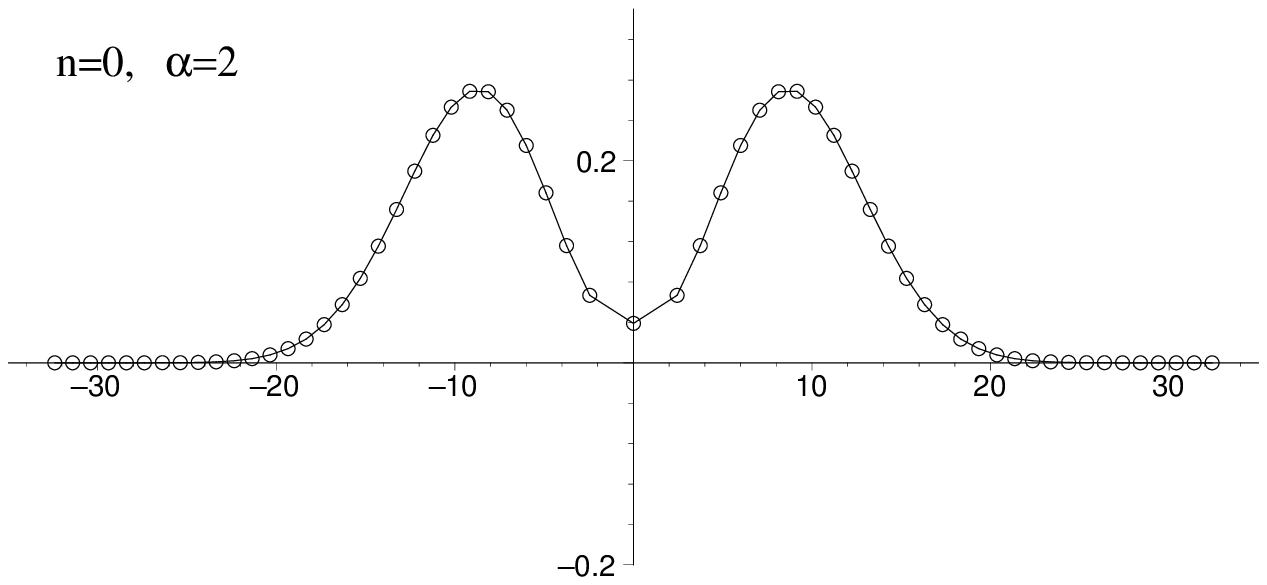} & \includegraphics[scale=0.55]{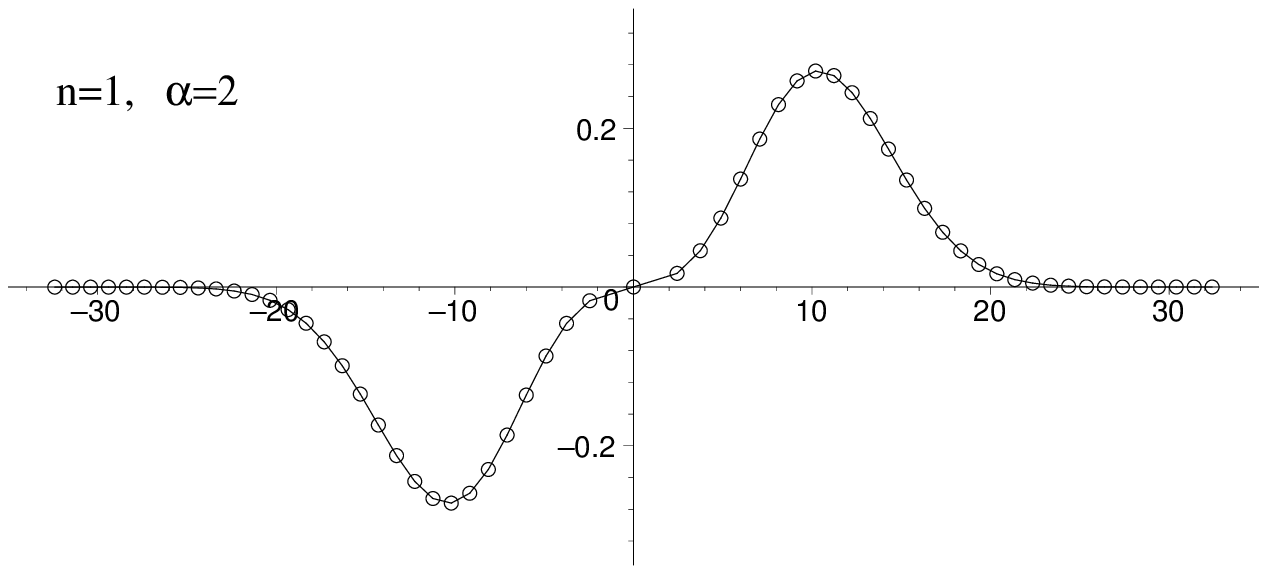} \\
\includegraphics[scale=0.55]{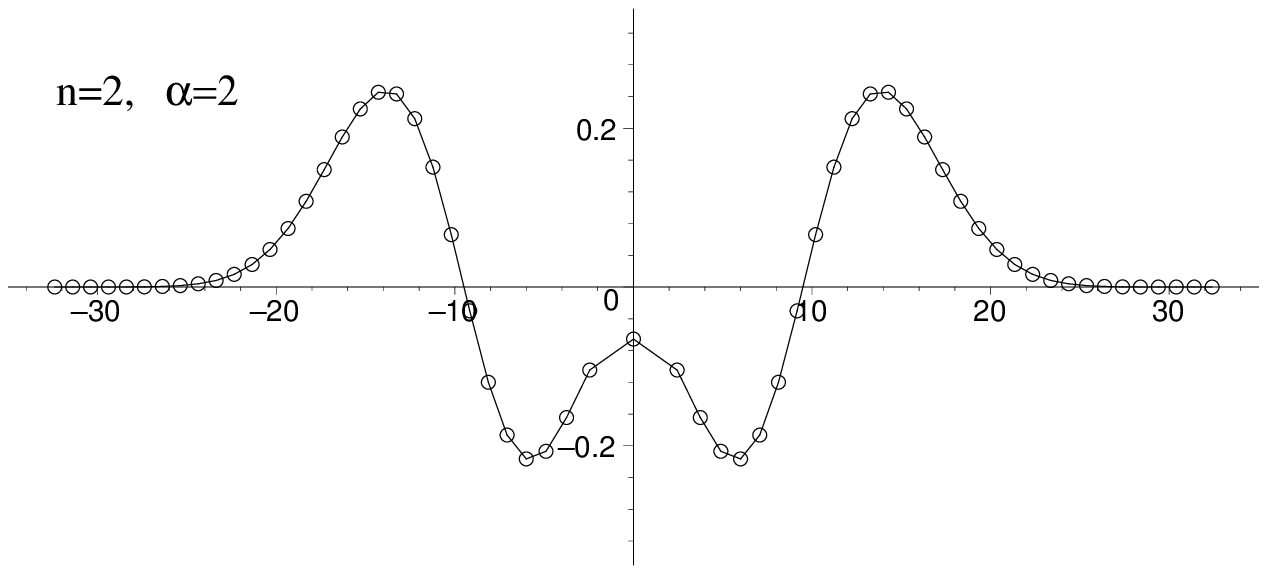} & \includegraphics[scale=0.55]{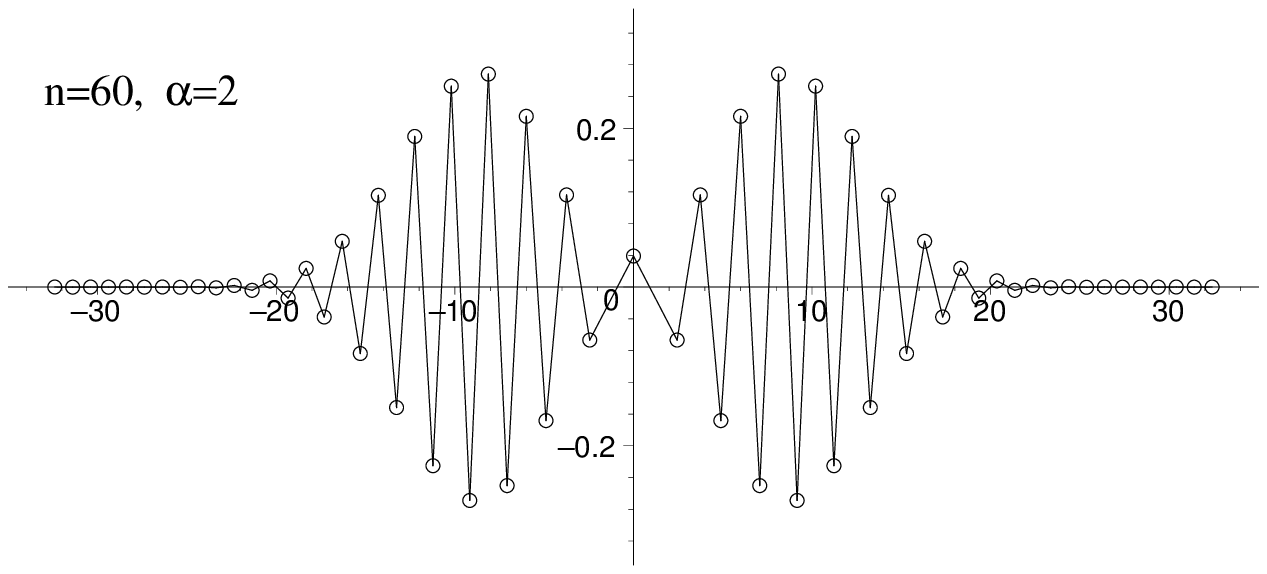} \\[-1mm]
\hline
\end{tabular} 
\caption{Plots of the discrete wavefunctions $\Phi^{(\alpha)}_n(q)$ in the representation with $j=30$. 
The four top figures are for $\alpha=-1/2$,
the middle figures for $\alpha=-0.7$, and the four bottom figures for $\alpha=2$. In each case, we plot
the wavefunctions for $n=0,1,2,60$.}
\label{fig2}
\end{figure}

Before studying some special properties of these position wavefunctions, and determining the momentum
wavefunctions, let us consider plots of these functions for some $\alpha$-values.
We choose a fixed value of $j$, say $j=30$, and plot some of the 
wavefunctions $\Phi^{(\alpha)}_n(q)$ for certain values of $\alpha$.
Since $\alpha=-\frac12$ is a special case (where $\su(2)_\alpha$ reduces to $\su(2)$), there are
three cases to be considered: $-1<\alpha<-\frac12$, $\alpha=-\frac12$ and $\alpha>-\frac12$.
In Figure~\ref{fig2} we take $\alpha=-\frac12$, $\alpha=-0.7$ and $\alpha=2$ respectively.
We also plot in each case the ground state $\Phi^{(\alpha)}_0(q)$, some low energy states $\Phi^{(\alpha)}_1(q)$ and $\Phi^{(\alpha)}_2(q)$,
and the highest energy state.
The plots are similar as in the $\uu(2)_\alpha$ case (where $j$ is half-integer). 
The most obvious difference is that 0 is part of the spectrum now. 
Another difference, but more difficult to see in the plots, is that the spectrum is not equidistant for
$\alpha\ne-\frac12$.
For $\alpha=-\frac12$, these plots coincide with the ones given in the $\su(2)$ model~\cite{Atak2005,Atak2001}.
The wavefunctions $\Phi^{(-1/2)}_n(q)$ are Krawtchouk functions. 
This follows also from the following transformations of ${}_3F_2$ series to ${}_2F_1$ series when $\alpha=-\frac12$,
according to~\cite[(48)]{Atak2005}
\begin{align}
& {\;}_3F_2 \left( \myatop{-k,k,-n}{1/2,-j} ; 1 \right)= (-1)^n \frac{\binom{2j}{2n}}{\binom{j}{n}} 
{\;}_2F_1 \left( \myatop{-2n,-j-k}{-2j} ; 2 \right),\\
& {\;}_3F_2 \left( \myatop{-k+1,k+1,-n}{3/2,-j+1} ; 1 \right)= -\frac{(-1)^n}{2k} \frac{\binom{2j}{2n+1}}{\binom{j-1}{n}} 
{\;}_2F_1 \left( \myatop{-2n-1,-j-k}{-2j} ; 2 \right). 
\end{align}

For $\alpha\ne-\frac12$, the plots are comparable with the parabose wavefunctions~\cite{JSV2011}.
One can indeed again study the behaviour of the discrete wavefunctions $\Phi^{(\alpha)}_n(q)$ when
the representation parameter $j$ tends to infinity.
In this process, one should pass from a discrete position variable $q$ to a continuous variable $x$.
This can be done by putting $q=j^{1/2} x$ and then compute the limit of $j^{1/4} \Phi^{(\alpha)}_n(q)$
for $j\rightarrow \infty$.
The actual computation is similar to the limit computation performed in~\cite[\S 4]{JSV2011}, so we shall not
give any details. 
Note that, due to~\eqref{Phi-even}, 
\begin{equation}
q^2 = (k+\alpha+\frac12)^2-(\alpha+\frac12)^2,\hbox{ or }
k=-\alpha-\frac12 \pm \sqrt{q^2+(\alpha+\frac12)^2}.
\label{kq}
\end{equation}
Using this last expression for $k$ in the ${}_3F_2$ expression of~\eqref{Phi-even}, 
and replacing herein $q$ by $j^{1/2} x$, the limit can be computed and yields:
\begin{equation}
\lim_{j\rightarrow\infty} j^{1/4} \Phi^{(\alpha)}_{2n}(j^{1/2} x ) = 
(-1)^n \sqrt{\frac{n!}{\Gamma(\alpha+n+1)}}\; |x|^{\alpha+1/2} e^{-x^2/2} L_n^{(\alpha)}(x^2),
\label{psi-even}
\end{equation}
in terms of Laguerre polynomials. Similarly, one finds from~\eqref{Phi-odd}:
\begin{equation}
\lim_{j\rightarrow\infty} j^{1/4} \Phi^{(\alpha)}_{2n+1}(j^{1/2} x ) = 
(-1)^n \sqrt{\frac{n!}{\Gamma(\alpha+n+2)}}\; x |x|^{\alpha+1/2} e^{-x^2/2} L_n^{(\alpha+1)}(x^2).
\label{psi-odd}
\end{equation}
The functions in the right hand side of~\eqref{psi-even}-\eqref{psi-odd} are known: they are the wavefunctions
of the parabose oscillator~\cite{Mukunda,Ohnuki,JSV2008} with parameter $a=\alpha+1>0$, see the appendix of~\cite{JSV2011}.

It remains here to consider the momentum wavefunctions.
Due to the fact that the spectrum of $\hat q$ and $\hat p$ is the same, and due to the similarity of 
the matrix of eigenvectors $V$ (compared to $U$), the expressions are analogous and we give only the final
result here:
\begin{align}
\Psi^{(\alpha)}_{2n} (p_{k}) & = -\frac{i}{\sqrt{2}} 
\sqrt{\frac{w(n;\alpha,\alpha,j)}{h(k;\alpha,\alpha,j)}}  
R_n \left( p_k^2; \alpha, \alpha, j\right), \qquad p_{k}^2 = k(2\alpha+k+1),
\label{Psi-even} \\
\Psi^{(\alpha)}_{2n+1} (p_{k}) & =
\sqrt{\frac{w(n;\alpha+1,\alpha+1,j-1)}{h(k-1;\alpha+1,\alpha+1,j-1)}} 
R_n \left( p_k^2-2(\alpha+1);\alpha+1,\alpha+1,j-1 \right).
\label{Psi-odd}
\end{align}

\section{The discrete Hahn-Fourier transform}

In canonical quantum mechanics, the momentum wavefunction (in $L^2(\R)$) is given by the Fourier transform of
the position wavefunction (and vice versa):
\[
\Psi(p)= \frac{1}{\sqrt{2\pi}} \int e^{-ipq}\Phi(q)dq.
\]
In the current case, we are dealing with discrete wavefunctions, and we should look for an analogue of this.
In terms of the notation of the previous section, let 
\begin{equation}
\Phi (q_k)=\left(
\begin{array}{c}
\Phi_0(q_k)  \\
\Phi_1 (q_k) \\
\vdots  \\
\Phi_{2j}(q_k)
\end{array}
\right), \qquad
\Psi (p_k)=\left(
\begin{array}{c}
\Psi_0(p_k)  \\
\Psi_1 (p_k) \\
\vdots  \\
\Psi_{2j}(p_k)
\end{array}
\right)\qquad (k=-j,\ldots,+j).
\label{PhiV-PsiV}
\end{equation}
So it is natural to define the discrete Fourier transform in this case as the matrix 
$F=(F_{lk})_{-j\leq l,k\leq +j}$ relating these two wavefunctions. In other words:
\begin{equation}
\Psi(p_l)=\sum_{k=-j}^j F_{kl}\;\Phi(q_k).
\label{F}
\end{equation}
As this generalized discrete Fourier transform maps Hahn polynomials into Hahn polynomials, we shall
refer to it as the discrete Hahn-Fourier transform.
By~\eqref{UPhi}-\eqref{VPsi}, the columns of $V$ consist of the column vectors $\Psi(p_k)$ ($k=-j..,+j$) 
and similarly for the matrix $U$. So~\eqref{F} actually means that $V = U F$, or:
\begin{equation}
F=U^T V.
\label{FVU}
\end{equation}
Using the explicit matrix elements from $U$ and $V$, \eqref{Uj}-\eqref{Uodd} and~\eqref{V},
this leads to following form of the matrix elements of $F$:
\begin{align}
&F_{j-k,j\mp l}=F_{j+k,j\pm l}=-\frac{i}{2}\sum_{n=0}^j(-1)^n \tilde Q_k(n;\alpha,\alpha,j) \tilde Q_l(n;\alpha,\alpha,j)
\label{F1}\\
&\pm \frac12 \sum_{n=0}^{j-1}(-1)^n \tilde Q_{k-1}(n;\alpha+1,\alpha+1,j-1) \tilde Q_{l-1}(n;\alpha+1,\alpha+1,j-1),
\quad k,l=1,\ldots,j;\nn\\
&F_{j\mp k,j}=F_{j,j\mp k}=-\frac{i}{\sqrt{2}}\sum_{n=0}^j(-1)^n \tilde Q_k(n;\alpha,\alpha,j) \tilde Q_0(n;\alpha,\alpha,j), \quad k=1,\ldots,j;\label{F2}\\
& F_{jj}=-i\sum_{n=0}^j(-1)^n \tilde Q_0^2(n;\alpha,\alpha,j). \label{F3}
\end{align} 

In the following section, we shall determine explicit expressions for the above matrix
elements. 
But before that, we can already summarize some properties of the discrete Hahn-Fourier transform matrix $F$.
\begin{prop}
The $(2j+1)\times(2j+1)$-matrix $F$ is symmetric, $F^T=F$, and unitary, $F^\dagger F=F F^\dagger =I$. 
Furthermore, it satisfies $F^4=I$, so its eigenvalues are $\pm 1, \pm i$.
A set of orthonormal eigenvectors of $F$ is given by the rows of $U$, determined in Proposition~\ref{propU}.
The multiplicity of the eigenvalues depends on the parity of $j$. 
When $j=2n$ is even, then the multiplicity of $-i,1,i,-1$ is $n+1,n,n,n$ respectively.
When $j=2n+1$ is odd, then the multiplicity of $-i,1,i,-1$ is $n+1,n+1,n+1,n$ respectively.
\end{prop}

\noindent {\bf Proof.}
The symmetry of $F$ is easily seen from the expressions~\eqref{F1}-\eqref{F3}.
The unitarity of $F$ follows from~\eqref{FVU}, the orthogonality of the real matrix $U$ and the unitarity of $V$.
Again using~\eqref{FVU} and the orthogonality of $U$, one finds $F^2=F^TF= V^TUU^TV=V^TV$.
But the explicit form of $V^TV$ is known, see~\eqref{VTV}. Since $(V^TV)^2=I$, the result $F^4=I$ follows. 
So the eigenvalues can only be $\pm 1, \pm i$.
Using again~\eqref{FVU} and~\eqref{VJU}, one sees that $F=U^TV=U^T{\cal J}U$,  or
\[
F U^T = U^T {\cal J}.
\]
In other words, the columns of $U^T$ (or the rows of $U$) form a set of orthonormal eigenvectors
of $F$, and the eigenvalues of $F$ are found in the diagonal matrix ${\cal J}$. From the
explicit form of ${\cal J}$, the statement of the multiplicities follows. \mybox

Note that these properties are similar to those of the common discrete Fourier transform~\cite{McClennan,Atak1997}.
For the case $\alpha=-\frac12$, the matrix $F$ was already studied in~\cite{Atak1997}.
In that special case, the position and momentum wavefunctions~\eqref{PhiV-PsiV} are in terms of
Krawtchouk polynomials, and the corresponding discrete Fourier transform can be referred to as
the discrete Krawtchouk-Fourier transform. The matrix elements of $F$ are in that special
case again Krawtchouk functions~\cite{Atak1997}.

One of our main results is the explicit computation of the elements of $F$ for general $\alpha$.
Note from~\eqref{F1}-\eqref{F3} that all these matrix elements are of the form
$\sum_{n=0}^j(-1)^n \tilde Q_k(n;\alpha,\alpha,j) \tilde Q_l(n;\alpha,\alpha,j)$,
with $0\leq k,l \leq j$.
So apart from a factor $(h(k;\alpha,\alpha,j)h(l;\alpha,\alpha,j))^{-1/2}$, this expression is equal to:
\begin{align}
S(k,l,\alpha,j) &= \sum_{n=0}^j (-1)^n w(n;\alpha,\alpha,j) Q_k(n;\alpha,\alpha,j)Q_l(n;\alpha,\alpha,j) \label{S1}\\
&= \sum_{n=0}^j (-1)^n \binom{\alpha+n}{n}\binom{\alpha+j-n}{j-n} Q_k(n;\alpha,\alpha,j)Q_l(n;\alpha,\alpha,j) \label{S2}\\
&= \sum_{n=0}^j (-1)^n \binom{\alpha+n}{n}\binom{\alpha+j-n}{j-n} R_n(\lambda(k);\alpha,\alpha,j)R_n(\lambda(l);\alpha,\alpha,j).
\label{S3}
\end{align}
In the form~\eqref{S1}, the right hand side is just like the orthogonality relation of Hahn polynomials, except
for the extra factor $(-1)^n$. In the form~\eqref{S3}, one can see that $S(k,l,\alpha,j)$ is a special case of 
the Poisson kernel (or a bilinear generating function) for dual Hahn polynomials:
\[
\sum_{n=0}^j t^n \binom{\alpha+n}{n}\binom{\beta+j-n}{j-n} R_n(\lambda(k);\alpha,\beta,j)R_n(\lambda(l);\alpha,\beta,j).
\]
However, as far as we know a closed form expression for this Poisson kernel is not known.
In fact, the best one can do is re-express the product $R_n(\lambda(k);\alpha,\beta,j)R_n(\lambda(l);\alpha,\beta,j)$
into a ${}_8F_7$ hypergeometric series (following the approach of~\cite[\S 8.3]{Gasper},
where the $q$-analogue is given). 
Here we will show that this Poisson kernel does have a simple expression when $\beta=\alpha$ and $t=-1$.

We shall prove the following result, yielding the explicit expression for the discrete Hahn-Fourier transform 
matrix $F$:
\begin{theo}
The special Poisson kernel for dual Hahn polynomials $S(k,l,\alpha,j)$, where $k$ and $l$ are integers
with $0\leq k,l \leq j$, satisfies
\begin{equation}
S(k,l,\alpha,j)=(-1)^{k+l+j} S(k,l,\alpha,j),
\label{S0}
\end{equation}
so it is 0 whenever $k+l+j$ is odd. For the 
other cases, it is given by:
\begin{align}
S(2K,2L,\alpha,2J)=& \frac{2^{2J}(\frac12)_{J-K}(\frac12)_{J-L}(\alpha+1)_J(\alpha+J+1)_K(\alpha+J+1)_L}{(2J)!(\frac12)_J} \nn\\
& \times {\ }_4F_3 \left( \myatop{-K,K+\alpha +\frac12,-L,L+\alpha+\frac12}{\alpha+J+1,\alpha +1,-J} ; 1 \right)\label{KL1} \\
S(2K+1,2L+1,\alpha,2J)=
 &\frac{2^{2J}(\frac12)_{J-K}(\frac12)_{J-L}(\alpha+1)_{J+1}(\alpha+J+2)_K(\alpha+J+2)_L}{(2J)!J(\frac12)_J} \nn\\
&\times {\ }_4F_3 \left( \myatop{-K,K+\alpha +\frac32,-L,L+\alpha+\frac32}{\alpha+J+2,\alpha +1,-J+1} ; 1 \right)\label{KL2} \\
S(2K,2L+1,\alpha,2J+1)=
 &\frac{2^{2J+1}(\frac12)_{J-K+1}(\frac12)_{J-L}(\alpha+1)_{J+1}(\alpha+J+2)_K(\alpha+J+2)_L}{(2J+1)!(\frac12)_{J+1}}\nn\\
&\times  {\ }_4F_3 \left( \myatop{-K,K+\alpha +\frac12,-L,L+\alpha+\frac32}{\alpha+J+2,\alpha +1,-J} ; 1 \right)\label{KL3}
\end{align} 
and finally $S(2K+1,2L,\alpha,2J+1)$ is given by replacing $K$ and $L$ in the right hand side of~\eqref{KL3}.
\label{maintheo}
\end{theo}
First of all, note that $S(k,l,\alpha,j)=0$ for $k+l+j$ odd implies that in expression~\eqref{F1} only one of the two
parts survive (either the real part or else the imaginary part). 
Together with Theorem~\ref{maintheo} this implies that each element $F_{lk}$ is, up to a factor, equal to a
terminating Saalsch\" utzian ${}_4F_3(1)$ series. 
In other words, up to a factor each $F_{lk}$ is also a Racah polynomial~\cite{Koekoek}.
The unitarity of the matrix $F$ does not lead to any new relations for Racah polynomials, but just
follows from their discrete orthogonality relations.
 
Before discussing the proof of Theorem~\ref{maintheo}, let us examine what happens to this discrete Hahn-Fourier transform
in the limit when $j\rightarrow\infty$. 
Suppose that $j=2J$ is even, and let us first consider the limit of the imaginary matrix elements of $F$.
When $k$ and $l$ are even ($j=2J$, $k=2K$ and $l=2L$), then according to~\eqref{F1}:
\[
F_{j+k,j\pm l} = -\frac{i}{2} (h(k;\alpha,\alpha,j)h(l;\alpha,\alpha,j))^{-1/2} S(2K,2L,\alpha,2J),
\]
with $S(2K,2L,\alpha,2J)$ given by~\eqref{KL1}. Just as for the limit computation of the wavefunctions, see~\eqref{kq},
it is necessary to make the replacements
\[
k=\sqrt{q_k^2+(\alpha+\frac12)^2}-(\alpha+\frac12),\qquad l=\sqrt{p_l^2+(\alpha+\frac12)^2}-(\alpha+\frac12),
\]
and put $q_k=j^{1/2}x$ and $p_l=j^{1/2}p$ in the above expression.
After doing this, the limit of the ${}_4F_3$ series appearing in $S(2K,2L,\alpha,2J)$ is fairly easy to determine
by termwise computation:
\[
\lim_{j\rightarrow \infty} {\ }_4F_3 \left( \myatop{-\frac{k}{2},\frac{k}{2}+\alpha +\frac12,-\frac{l}{2},
\frac{l}{2}+\alpha+\frac12}{\alpha+\frac{j}{2}+1,\alpha +1,-\frac{j}{2} } ; 1 \right) =
{\;}_0F_1\left( \myatop{ - }{\alpha+1}; -\frac{1}{4}x^2p^2\right).
\]
The last expression is proportional to $J_\alpha(xp)$, where $J_\alpha$ is the Bessel function of the first kind~\cite{Temme}.
The limit of the factors in front of the ${}_4F_3$ series are elementary but a bit more tedious to compute, and we shall
not give the details of this computation.
Similarly, one has to determine the limit of the real matrix elements of $F$. Adding both contributions,
one finds:
\[
\lim_{j\rightarrow \infty} j^{1/2} F_{j+k,j\pm l} = \frac12 \left(
|xp|^{1/2} J_\alpha(|xp|) + ixp |xp|^{-1/2} J_{\alpha+1}(|xp|)
\right).
\]
The function in the right hand side is known: it is the kernel of the generalized Fourier transform,
studied by Mukunda {\em et al} in the context of the parabose oscillator~\cite{Mukunda}.
So one can conclude that our discrete Hahn-Fourier transform tends to the generalized Fourier transform
of~\cite{Mukunda} for large values of $j$.

\section{Computation of the discrete Hahn-Fourier transform matrix elements}

The purpose of this section is the proof of Theorem~\ref{maintheo}.
In fact, we shall show~\eqref{S0} and~\eqref{KL1}; the proof of~\eqref{KL2} and~\eqref{KL3} is analogous.

Let us first collect some known transformation formulas for hypergeometric series.
The first is Thomae's transformation of a terminating Saalsch\"utzian ${}_4F_3(1)$ series~\cite[(2.4.1.7)]{Slater}, 
\cite[(7.2.1)]{Bailey}:
\begin{equation}
{\;}_4F_3 \left( \myatop{a,b,c,-N}{e,f,g} ; 1 \right)=
\frac{(f-c)_N(e+f-a-b)_N}{(f)_N(e+f-a-b-c)_N} {\;}_4F_3 \left( \myatop{e-a,e-b,c,-N}{e,e+f-a-b,e+g-a-b} ; 1 \right)
\label{Thomae}
\end{equation}
where $e+f+g=1+a+b+c-N$.
For the Hahn polynomials appearing in~\eqref{S1}, the following transformation formula holds~\cite[p.~186]{Gasper1974}:
\begin{equation}
Q_k(x;\alpha,\alpha,N)= (-1)^k Q_k(N-x;\alpha,\alpha,N).
\label{Q-symm}
\end{equation}
In the same paper~\cite[(3.18)-(3.19)]{Gasper1974}, one can find the following ${}_4F_3$ expressions 
for Hahn polynomials with parameter $\beta=\alpha$:
\begin{align}
Q_{2k}(x;\alpha,\alpha,N)&= 
{\;}_4F_3 \left( \myatop{-k,k+\alpha+\frac12,-x,x-N}{\alpha+1,-N/2,(1-N)/2} ; 1 \right), \label{Qeven}\\
Q_{2k+1}(x;\alpha,\alpha,N)&= \frac{N-2x}{N}
{\;}_4F_3 \left( \myatop{-k,k+\alpha+\frac32,-x,x-N}{\alpha+1,(1-N)/2,(2-N)/2} ; 1 \right). \label{Qodd}
\end{align}

Next, we need an equation that goes back to a classical identity for $6j$-coefficients.
For $6j$-coefficients of $\su(2)$, the following holds~\cite[(9.8.4)]{Varshalovich}:
\begin{equation}
\sum_x (-1)^{p+q+x}(2x+1) 
\sixj{a}{b}{x}{c}{d}{p} \sixj{a}{b}{x}{d}{c}{q} = \sixj{a}{c}{q}{b}{d}{p}. 
\label{su2-6j}
\end{equation}
It turns out that we need the corresponding identity for $6j$-coefficients of positive discrete series representations
of $\su(1,1)$, which reads:
\begin{equation}
\sum_{j_{23}} (-1)^{j_{13}+j_{23}-j'}\; U^{k_1,k_2,k_{12}}_{k_3,k,k_{23}} U^{k_1,k_3,k_{13}}_{k_2,k,k_{23}} =
U^{k_2,k_1,k_{12}}_{k_3,k,k_{13}}. 
\label{su11-6j}
\end{equation}
Herein, $U^{k_1,k_2,k_{12}}_{k_3,k,k_{23}}$ is the standard notation for the 
Racah coefficient of $\su(1,1)$~\cite{LNM}. In~\eqref{su11-6j}, $k_1$, $k_2$ and $k_3$ are $\su(1,1)$
representation labels (i.e.\ they are positive real numbers), and in a common notation~\cite{LNM}
\[
k_{12}=k_1+k_2+j_{12}, \quad k=k_{12}+k+3+j', \quad k_{13}=k_1+k_3+j_{13},\quad k_{23}=k_2+k_3+j_{23},
\]
with $j_{12}$, $j'$, $j_{13}$ and $j_{23}$ all nonnegative integers. 
The summation in the right hand side of~\eqref{su11-6j} runs over all $j_{23}$ with $0\leq j_{23} \leq j'+j_{12}$.
Using the explicit expression in terms of a ${}_4F_3(1)$ series for the Racah coefficients in~\eqref{su11-6j}, given 
by~\cite[(4.41)]{LNM}, and making the following replacements:
\[
(j',j_{12},j_{13}) \longrightarrow (p,q,r),\qquad 
(\frac{k_1}{2}, \frac{k_2}{2},\frac{k_3}{2}) \longrightarrow (a,b,c), \qquad
j_{23} \longrightarrow n,
\]
one arrives at the following identity between terminating Saalsch\"utzian ${}_4F_3(1)$ series:
\begin{align}
& \sum_{n=0}^{p+q} \binom{p+q}{n} \frac{(b+c+2n-1)}{(b+c+n-1)}
 \frac{(b+c)_n(a+b+c+p+q-1)_n}{(b+c+p+q)_n(1-a-p-q)_n} \nn \\
& \times  {\;}_4F_3 \left( \myatop{-n,n+b+c-1,-q,q+a+b-1}{b,a+b+c+p+q-1,-p-q} ; 1 \right)
{\;}_4F_3 \left( \myatop{-n,n+b+c-1,-r,r+a+c-1}{c,a+b+c+p+q-1,-p-q} ; 1 \right) \nn\\
&= (-1)^{p-r} \frac{(b+c)_{p+q}(a)_r(a)_q}{(a)_{p+q}(b)_q(c)_r}
{\;}_4F_3 \left( \myatop{-q,q+a+b-1,-r,r+a+c-1}{a,a+b+c+p+q-1,-p-q} ; 1 \right).
\label{id1}
\end{align}

This is not yet in the form needed for our proof of Theorem~\ref{maintheo}.
But after performing Thomae's transformation~\eqref{Thomae} on the 2nd and 3rd ${}_4F_3(1)$ series
appearing in~\eqref{id1}, and replacing $r$ by $p+q-r$, we have the following:
\begin{lemm}
Let $p$, $q$ and $r$ be nonnegative integers, and $a$, $b$ and $c$ arbitrary parameters, then
\begin{align}
& \sum_{n=0}^{p+q} \binom{p+q}{n} \frac{(b+c+2n-1)}{(b+c+n-1)}
 \frac{(b+c)_n(b)_n}{(c)_n(b+c+p+q)_n} \nn \\
& \times  {\;}_4F_3 \left( \myatop{-n,n+b+c-1,-q,q+a+b-1}{b,a+b+c+p+q-1,-p-q} ; 1 \right)\nn\\
&\times {\;}_4F_3 \left( \myatop{-n,n+b+c-1,-r,r+1-a-c-2p-2q}{b,1-a-p-q,-p-q} ; 1 \right) \nn\\
= &(-1)^{r} \frac{(b+c)_{p+q}(1-c-p-q)_r}{(c)_p(a+b+c+p+q-1)_{q}(1-a-p-q)_r} \nn\\
&\times {\;}_4F_3 \left( \myatop{-q,q+a+b-1,-r,r+1-a-c-2p-2q}{b,1-c-p-q,-p-q} ; 1 \right).
\label{id2}
\end{align}
\end{lemm}

Note that the identities~\eqref{id1}-\eqref{id2} show at first sight some similarity with the expansion formulas
constructed in~\cite{Lievens} (or some $q$-analogues in~\cite{Gasper2000}); they turn out to be quite different however.

We now come to the final part of this section:

\noindent
{\bf Proof of Theorem~\ref{maintheo}.}
First of all, from~\eqref{S1}, note the symmetry $S(l,k,\alpha,j)=S(k,l,\alpha,j)$.
Starting from~\eqref{S2}, using~\eqref{Q-symm} for $Q_k$ and $Q_l$, and then reversing the order of summation
(i.e.\ replace $n$ by $n-j$), one finds~\eqref{S0}.
This implies that $S(k,l,\alpha,j)=0$ whenever $k+l+j$ is odd.
So we need to simplify the expression only when $k+l+j$ is even.
We shall do this explicitly in the case that $k$, $l$ and $j$ are even: $k=2K$, $l=2L$ and $j=2J$, 
with $K$, $L$ and $J$ nonnegative integers (the other three cases are similar).
So we need to compute
\begin{equation}
S(2K,2L,\alpha,2J) 
= \sum_{n=0}^{2J} (-1)^n \binom{\alpha+n}{n}\binom{\alpha+2J-n}{2J-n} Q_{2K}(n;\alpha,\alpha,2J)Q_{2L}(n;\alpha,\alpha,2J).
\end{equation}
Using~\eqref{Qeven}, this can be written as 
$S(2K,2L,\alpha,2J) = \sum_{n=0}^{2J} T_n(K,L)$, where
\begin{align}
T_n(K,L)& = (-1)^n \binom{\alpha+n}{n}\binom{\alpha+2J-n}{2J-n}
{\;}_4F_3 \left( \myatop{-K,K+\alpha+\frac12,-n,n-2J}{\alpha+1,-J,-J+\frac12} ; 1 \right)\nn\\
& \qquad\times{\;}_4F_3 \left( \myatop{-L,L+\alpha+\frac12,-n,n-2J}{\alpha+1,-J,-J+\frac12} ; 1 \right).
\label{TnKL}
\end{align}
Making an appropriate replacement of the summation variable, it is easy to see that
\[
\sum_{n=J}^{2J} T_n(K,L) = \sum_{n=0}^J T_n(K,L),
\]
hence we can split the total sum $\sum_{n=0}^{2J} T_n(K,L)$ in two equal parts; taking care of the overlapping middle term,
there comes
\begin{equation}
S(2K,2L,\alpha,2J) = 2 \left(\sum_{n=0}^{J-1} T_n(K,L) +\frac12 T_J(K,L) \right).
\label{S2S}
\end{equation}
In order to perform the summation in the right hand side of~\eqref{S2S}, one can make the following
substitution in~\eqref{id2}:
\begin{equation}
p=J-K,\ q=K,\ r=L,\ a=\frac12,\ b=\alpha+t,\ c=-2J-\alpha, \label{subs}
\end{equation}
and then take the limit $t\rightarrow 1$ (in fact, for all terms in~\eqref{id2} one can immediately take $t=1$, except
for the term with $n=p+q=J$ the limit process is actually necessary). 
Consider, after the substitution~\eqref{subs}, the $n$th term in the left hand side of~\eqref{id2}, with $0\leq n<p+q=J$:
the coefficient in front of the two ${}_4F_3(1)$'s becomes:
\[
\frac{(2J)!}{(\alpha+1)_{2J}} (-1)^n \binom{\alpha+n}{n}\binom{\alpha+2J-n}{2J-n}.
\]
Consider similarly the term with $n=p+q=J$; here the coefficient becomes, when $t\rightarrow 1$:
\[
\frac12 \frac{(2J)!}{(\alpha+1)_{2J}} (-1)^J \binom{\alpha+J}{J}\binom{\alpha+J}{J}.
\]
Furthermore, it is easy to see that under this substitution~\eqref{subs} and $t\rightarrow 1$ the two ${}_4F_3(1)$'s
in the left hand side of~\eqref{id2} become the ${}_4F_3(1)$ expressions of~\eqref{TnKL}.
Thus, up to the coefficient $\frac{(2J)!}{(\alpha+1)_{2J}}$, the right hand side of~\eqref{S2S} can be summed
according to~\eqref{id2}, and one finds, after simplifications:
\begin{align*}
S(2K,2L,\alpha,2J)& = \frac{2^{2J}(\frac12)_{J-K}(\frac12)_{J-L}(\alpha+1)_J(\alpha+J+1)_K(\alpha+J+1)_L}{(2J)!(\frac12)_J} \\
& \times {\ }_4F_3 \left( \myatop{-K,K+\alpha +\frac12,-L,L+\alpha+\frac12}{\alpha+J+1,\alpha +1,-J} ; 1 \right),
\end{align*}
proving~\eqref{KL1}. \mybox

\section{Summary}

The most popular finite oscillator model, especially for applications in quantum optics, 
is based on the Lie algebra $\su(2)$ or $\so(3)$~\cite{Atak2005,Atak2001,Atak2001b}.
The dimension of the model depends on the $\su(2)$ representation label~$j$, which is
an integer or half-integer positive number. 
Its mathematical properties have been studied well, in particular the properties
of the wavefunctions given by Krawtchouk functions.
These wavefunctions have interesting plots, and in the limit $j\rightarrow \infty$ these
discrete wavefunctions tend to the continuous canonical oscillator wavefunctions.
The discrete Fourier transform turning position wavefunctions into momentum wavefunctions
has also been investigated in this case~\cite{Atak1997}.

A first type of deformation of this $\su(2)$ model was offered by its $q$-deformation.
The $su_q(2)$ model for the finite oscillator was investigated in~\cite{Ballesteros,AKW}.
The main properties are: the position operator has a discrete anharmonic spectrum, and the wavefunctions
are given in terms of dual $q$-Krawtchouk polynomials~\cite{AKW}.

We have now considered two different type of deformations of the $\su(2)$ model, closely related to each other.
The first deformation $\uu(2)_\alpha$ was given in~\cite{JSV2011} and allows a deformation of the even-dimensional
representations only ($j$ half-integer). The second deformation $\su(2)_\alpha$ was the topic of this paper, and 
allows a deformation of the odd-dimensional representations only ($j$ integer).
Both cases have a deformation parameter $\alpha>-1$ and for $\alpha=-\frac12$ they reduce to the
nondeformed $\su(2)$ model.
The (discrete) spectrum of the position operator can be constructed explicitly in the deformed algebras.
In the case of $\uu(2)_\alpha$ it is equidistant except for an extra shift in the middle of the spectrum.
In the case of $\su(2)_\alpha$ it is no longer equidistant but has the simple form $\pm \sqrt{k(k+2\alpha+1)}$
($k=0,1,\ldots,j$). For $k$ sufficiently large, this behaves like $\pm(k+\alpha+\frac12)$, so it tends to
an equidistant distribution for large $k$.

The position and momentum wavefunctions have been constructed explicitly for the deformed models.
They are given in terms of Hahn polynomials.
For $\uu(2)_\alpha$, the even wavefunctions are normalized Hahn polynomials with parameters $(\alpha,\alpha+1)$,
and the odd wavefunctions with parameters $(\alpha+1,\alpha)$. For $\su(2)_\alpha$,
the even wavefunctions are normalized Hahn polynomials with parameters $(\alpha,\alpha)$, 
and the odd wavefunctions with parameters $(\alpha+1,\alpha+1)$.
The plots of these discrete wavefunctions have nice properties, and in both cases they tend
to the parabose wavefunctions when $j$ is large. 
For $\alpha\rightarrow -\frac12$ they tend to the Krawtchouk wavefunctions of the $\su(2)$ model;
and of course for $\alpha\rightarrow -\frac12$ and $j\rightarrow \infty$ they tend to the 
canonical oscillator wavefunctions in terms of Hermite polynomials.

An interesting extra object studied in this paper is the operation that transforms position wavefunctions into
momentum wavefunctions, i.e.\ the discrete analogue of the Fourier transform.
For the case of $\su(2)_\alpha$, this discrete Hahn-Fourier transform has been constructed explicitly,
and is determined by a matrix $F$. 
This matrix shares many classical properties with the common discrete Fourier transform matrix.
In fact, it has the extra interesting feature that there is a natural basis of eigenvectors of $F$.
The main computational result of the paper is the proof that the matrix elements of $F$ have a simple
form in terms of terminating Saalsch\"utzian ${\,}_4F_3(1)$ series, i.e.\ in terms of Racah polynomials.
Note that for $\uu(2)_\alpha$ this Hahn-Fourier matrix was not determined in~\cite{JSV2011}, but 
knowing the results for $\su(2)_\alpha$ it should be a routine computation to do this.

\section*{Acknowledgments}
E.I.~Jafarov was supported by a postdoc fellowship from the Azerbaijan National Academy of Sciences.
N.I.~Stoilova was supported by project P6/02 of the Interuniversity Attraction Poles Programme (Belgian State -- 
Belgian Science Policy).

\newpage

\newpage

\end{document}